\documentclass[12pt]{article}

\usepackage{newtxtext,newtxmath}

\usepackage{graphicx}

\usepackage[letterpaper,margin=1in]{geometry}

\renewenvironment{abstract}
	{\quotation}
	{\endquotation}

\date{}

\makeatletter
\renewcommand{\fnum@figure}{\textbf{Figure \thefigure}}
\renewcommand{\fnum@table}{\textbf{Table \thetable}}
\makeatother

\usepackage{scicite}

\usepackage{url}

\usepackage{bm}

\def\scititle{
	Snap and Jump: How Elastic Shells Pop Out
}
\title{\bfseries \boldmath \scititle}

\author{
	Takara Abe$^{1}$,
	Isamu Hashiguchi$^{2}$, 
	Yukitake Nakahara$^{1}$,
    \and
    Shunsuke Kobayashi$^{2}$, 
    Ryuichi Tarumi$^{2}$, 
    Hidetoshi Takahashi$^{1,3}$,
    \and
    Genya Ishigami$^{1,3}$,
    Tomohiko G. Sano$^{1,3*}$
    \and
	\small${}^{1}$School of Integrated Design Engineering, Graduate School of Science and Technology,
    \and 
    \small Keio University, 3-14-1 Hiyoshi, Yokohama, Kanagawa, 2238522, Japan\and
	\small${}^{2}$Graduate School of Engineering Science, Osaka University, \and
    \small 1-3 Machikaneyama, Toyonaka, Osaka, 5608531, Japan
    \and
    \small${}^{3}$Department of Mechanical Engineering, Faculty of Science and Technology, 
    \and
    \small Keio University, 3-14-1 Hiyoshi, Yokohama, Kanagawa, 2238522, Japan
    \and
	\small{$^*${Corresponding author. Email: sano@mech.keio.ac.jp}}\and
}

\begin{document} 

\maketitle


\begin{abstract} \bfseries \boldmath
Grip, walk, crawl, and jump. 
Soft robots are integrated functional structures composed of compliant mechanisms, whose activity spans various industrial applications such as surgery, healthcare, surveillance, and even planetary exploration. 
One of their promising mobility mechanism is snap-buckling; the instability mode of flexible structures passing from one equilibrium state to another can instantaneously generate large power for its motion. Predicting their performance with even simple geometry requires disentangling material, geometric nonlinearity, and contact, thereby still being an open challenge to date. 
Here, we study the jumping dynamics of hemispherical elastic shells driven by snap-buckling, as a model system of soft jumping mechanisms, combining experiments, simulations, and analytical theory. 
We find that the contact transition dynamics trigger the jumping phenomenon upon snap-buckling by constructing the analytical predictions with shell elasticity in excellent agreement with both experiments and simulations.  
Despite the simple geometry of the shell, its dynamical performance primarily relies on a complex interplay between elasticity, geometry, and contact friction. 
By elucidating the dynamics of the building blocks of soft robots that undergo large deformations, we can build their predictive experimental and numerical framework. Our research paves the way for designing soft robots suitable for the required loading conditions or structural requirements without empirical methods.
\end{abstract}

\noindent
Soft robots are highly adaptable in shape, allowing continuous and flexible deformation~\cite{Rus2015, Runciman2019, Chung2019, DEVAUCORBEIL2020185, Yang2021, Aubin2022, Park2023, Qin2023}.
They often consist of slender and flexible structures such as rods, plates, and shells that can fit and adapt to surrounding structures or environments through their continuous degree of freedom. 
Combining the energy inputs, such as pneumatic~\cite{Shepherd2011, Overvelde2015}, fluids~\cite{Abdullah2018}, or material nonlinearities~\cite{An2012, Haines2014, Lum2016, Wang2020, Minori2022, Sano2022_1, Sano2022_2, Hebner2023}, with the elastic instability of structures (\textit{e.g.,} snap-through buckling), one can realize fast and predictive deformation and motions~\cite{Rothemund2018, Sano2019, Gorissen2020, Vasios2021, Cao2021, Zhang2022, Abbasi2023, Wang2023, Tang2024, Li2024, Ching2024}.
For example, inspired by the children's popping toys (jumping poppers), pneumatic elastomeric actuators made of two spherical shells have been developed, whose energy mechanism has been studied in detail{~\cite{Gorissen2020}}.

Snap-through or snap-buckling of slender structures (\textit{e.g.,} pressurized thin spherical shell) is a canonical pathway for soft robotic systems.
As the buckling instability results from a complex interplay of boundary conditions, material property, geometric nonlinearity, and contact mechanics~\cite{Zoelly1915, Biezeno1935, Updike1970, Audoly2010, Nasto2013, Lopez2017, Taffetani2018, Zhao2022, Sahli2024}, the prediction of their mechanical performance is still a challenging problem to date~\cite{Bazant2010}.
Hence, building predictive frameworks for soft robots, where slender structures are assembled in a complex manner, requires extensive research effort from experimental, theoretical, and computational perspectives. 
In other words, for a given material and structural design, it is highly nontrivial to predict the performance and characteristics of soft robots or actuators without systematic parameter surveys.
Unraveling the multibody dynamics of building blocks of existing soft robot designs would accommodate insight into optimizing their ability and performance.

Here, we investigate the fast snapping dynamics of elastic spherical shells placed on a rigid plane to uncover the jumping mechanism of spherical shells as a building block of jumping soft robots (Fig.~\ref{fig:1}(a-c)).
We build the predictive framework for the jumping performance of elastic spherical shells, i.e., we construct the formulas for the characteristics of the jumping process and the maximum jumping height.
Our shells of the radius of curvature $R$, thickness $h$, and polar angle $\varphi(= 80^{\circ})$ are pneumatically controlled and are constrained one-dimensionally, translating vertically to precisely quantify the vertical catapulting motion (Fig.~\ref{fig:1}).
In parallel, we perform numerical simulations based on the material point method (MPM), which is an efficient computational method for contact dynamics with geometric nonlinearity~\cite{Sulsky1995, Jiang2015, Hu2019, DEVAUCORBEIL2020185, Qin2023}.
Combining experiments, simulations, and theory, we find that the elasticity and geometry of shells can predict jumping performances, such as contact dynamics, jumping force, and maximum height. 
Our proposed framework for elastic shells would be applicable to estimate existing or novel jumping soft robots without empirical studies through parametric surveys, complementing the recent pioneering work of inflatable jumpers{~\cite{Gorissen2020}}.


\subsection*{Overview of jumping experiments}

\begin{figure}[h]
    \centering
    \includegraphics[width =0.6\textwidth]{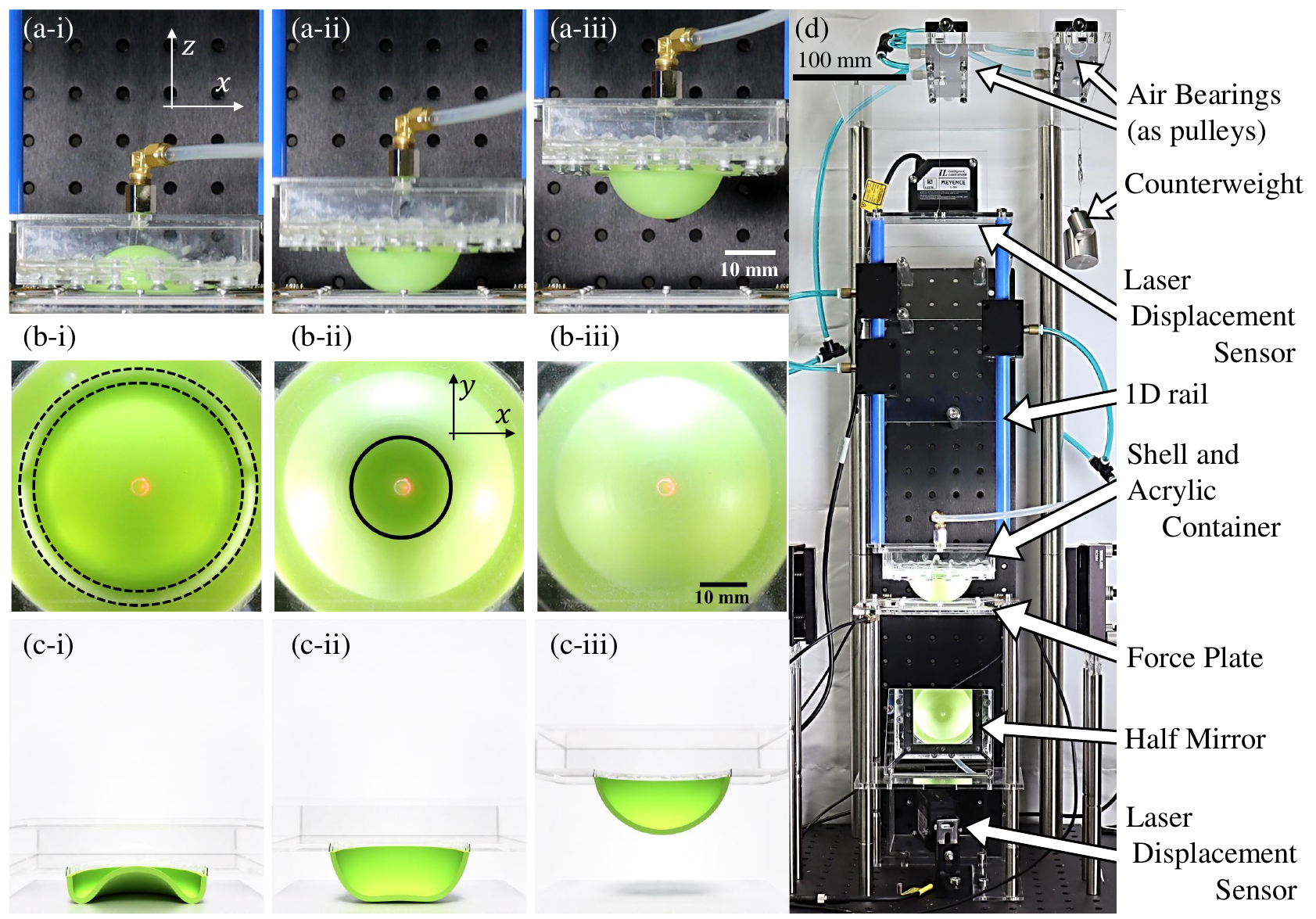}
    \caption{\textbf{Problem definition. The pneumatically-controlled elastic shell jumps on a rigid substrate.}
    (a-i)-(a-iii)~The front view of the typical jumping process of our elastic spherical shell ($(R, h)= (25, 2.0)$~mm). The shell is attached to the pneumatically controlled container. (a-i) We depressurize the container and place the buckled shell on the substrate. After recovering nearly-atmospheric pressure for the container instantly, (a-ii) the shell restores its original shape, and (a-iii) is catapulted by impacting the substrate.
    (b-i)-(b-iii)~The bottom view of the jumping process of (a-i)-(a-iii).
    (b-i) The buckled shell is initially set to contact the substrate via a ring-like shape (highlighted by the region surrounded by two dashed circles). (b-ii)~As the shell is restored, the apex of the shell contacts the substrate, forming disc-like contact, and (b-iii) then the shell snaps and jumps from the substrate.
    (c-i)-(c-iii)~The snapshots of the cross-sectional view of the simulated shell in a jumping process ($(R, h)= (25, 2.0)$~mm).
    (d)~Photograph of the experimental apparatus. The set of the container and shell is connected with the one-dimensional rail and counterweight via air bearings. In each jumping test, the vertical position and pressure of the container, contact radius and apex displacement of the shell, and normal reaction force of the substrate are measured simultaneously by combining the pressure sensor, two laser displacement sensors, the half mirror, the high-speed camera, and the custom-made force sensor plate.}
    \label{fig:1}
\end{figure}

Our thin elastic shell (Fig.~\ref{fig:2}(d) inset) is glued to an acrylic container whose internal pressure is controlled by a syringe and valve. The set of shell and acrylic container, which we call \textit{popper}, is connected to two parallel shafts (1D rail) sliding along air bearings (13mm ID Air Bushing, Newway, USA), allowing the popper to jump vertically with nearly-zero friction. The popper is coupled with a Nitinol wire to a counterweight via pulleys to control the net weight, $mg$ (Fig.~\ref{fig:1}(d)).
We measure the vertical location of the popper, $z$, the contact radius between the shell and substrate, $r$, the distance between the shell apex and substrate, $w$, the reaction force of the substrate, $F$, and the inner pressure of the container $p$, simultaneously. With the aid of the synchronized data, we detail the mechanisms of the jumping process below.

The experimental sequence is as follows (Fig.~\ref{fig:1}(a)). When the air pressure in the container $p$ is lowered, the shell is subjected to a pressure difference from the atmosphere, $\delta p$. When the pressure difference, $\delta p$, exceeds the critical value, the shell buckles. We then ground it to the substrate (the force plate) (Fig.~\ref{fig:1}(a-i)) and open the valve so that the inner pressure instantaneously reaches the nearly atmospheric pressure. The shell undergoes snap-through instability (to recover the natural shape) (Fig.~\ref{fig:1}(a-ii)) and impacts the substrate to jump (Fig.~\ref{fig:1}(a-iii)).

\subsection*{Jumping process of spherical shells}

The typical time series data of our experiment and simulation ($(R, h)$ = (30, 2.0) mm, $mg=$0.48~N) are shown in Fig.~\ref{fig:2}(a). 
When the valve is opened at time $t=0$, the pressure difference applied to the shell $\delta p$ drops from the initial value, $\delta p_j$, with the rapid air inflow (Fig.~\ref{fig:2}(a-iv)), and the shell starts recovering the original shape (seen as the decrease of $w$ and $r$ in Fig.~\ref{fig:2}(a-ii)). 
The pressure difference at $t$ = 0, $\delta p_j$, is related to the initial buckling amplitude of the shell for each experiment (Fig.~\ref{fig:2}(a-iv)).
As the contact radius decreases, the popper lifts ($z$ increases), with the rim of the buckled shell being in contact with the substrate (Fig.~\ref{fig:2}(a-i)). 
{The contact force $F$ gradually increases with the damped oscillation as the popper rises.}
The damped oscillation of $F$ at $t\lesssim0.4~{\rm seconds}$ is due to a longitudinal wave of the wire connecting the popper and the counterweight. 

The shell snaps at $t\simeq0.6~{\rm s}$ highlighted by the abrupt drop of $w$ toward $w\simeq 0$ and the sharp peak of the contact force $F^*$, where the outer surface of the shell is in areal contact with the substrate. 
Prior to the snap-buckling, only the rim of the indented shell touches the substrate, and the contact force is applied via a ring-like contact shape. At the snap onset, when the contact radius reaches the critical value $r^*$, the inside of the rim contacts with the substrate by area, where the contact geometry of the shell transits from a ring-like to a disk-like shape. The contact transition here generates a huge impulsive force $F^*$, and the popper leaves the substrate within the interval of $t^*$. Thereafter, the height of the popper, $z$, reaches the maximum height, $H$.

\subsection*{Contact transition on a rigid substrate}\label{sec:ContactTrans}

We here employ the scaling analysis for the characteristic quantities, $r^*, F^*, t^*$, and the apex displacement at the snap onset, $e^*$, based on the previous study for the deformation of spherical shells pushed by a plane \cite{Pauchard1997, Nowinka1994, Updike1972}.
Suppose that the shell is pushed by a plane by the height $e$ and that the apex of the shell is flattened, forming a disk-like contact shape of radius $r\sim\sqrt{e R}$. The material length of the indented region is shortened by $\sim r^3/R^2$ (from $R\sin^{-1}(r/R)$ to $r$), hence the strain is $\epsilon\sim(r/R)^2$. Balancing the elastic bending, $Eh^3/R^2$, and stretching energy per unit area, $Eh\epsilon^2$, we obtain the characteristic disk contact radius and apex displacement as $r_c = \sqrt{hR}$ and $e_c = h$, respectively. The stored elastic energy corresponds to the work done by the external force, $Fe$, giving the characteristic force as $F_c = Eh^3/R$~\cite{Landau1980, Pogorelov1988}. 
We expect that the critical radius, $r^*$, apex displacement, $e^*$, and force $F^*$ at the jump onset obey the same scaling law as $r^*\sim r_c$, $e^*\sim e_c$ and $F^*\sim F_c$, respectively. 
By introducing the corresponding dimensionless prefactors, $\alpha_r$, $\alpha_e$, and $\alpha_F$, we obtain
\begin{eqnarray}
    r^* = \alpha_r r_c,~~e^* = \alpha_e h,~~F^* = \alpha_F F_c\label{eq:rf}.
\end{eqnarray}
The precise values of prefactors would be determined analytically through the stability analysis of the ring-contact~\cite{Audoly2010}. Nevertheless, we can set their upper and lower bounds as $1.2<\alpha_r<1.6$, $1.8<\alpha_e<4.5$, and $5.3<\alpha_F<8.5$, where ring and disk contact states are bistable for the given loading force (See supplementary information). We plot the upper and lower bounds of the predictions Eq.~(\ref{eq:rf}) as the boundaries of the shaded regions in Fig.~\ref{fig:2}(b)-(d).
Our experimental and simulation results sit within our theoretical prediction without fitting parameters, indicating that the contact-geometry transition upon snap-buckling drives the jumping process.

The jumping dynamics follows balancing the inertia of the popper $\sim me/t_c^2$ and the flattening force $F^*$. The characteristic contact time scale, $t_c$, is then predicted as $t_c=\sqrt{mR/Eh^2}$, which agrees with experiments and simulations as $t^*=\alpha_t t_c$ with the prefactor $\alpha_t$ satisfying $\alpha_t = \sqrt{3\alpha_e/\alpha_F}$, despite the complexity of the overall experimental system (Fig.~\ref{fig:2}(d)). Note that our predictions have no adjustable parameters except the ambiguity of the prefactors bounded by analytical theory. 
We have revealed that the characteristic quantities in the jumping process obey the formulas based on the contact transition of the shell pushed by a plane without any adjustable parameters. 
In other words, we can conclude that the jumping dynamics of the popper is triggered by the contact-geometry transition of the shell.

\subsection*{Maximum jumping height}

We have established the framework for the characteristic quantities in the jumping process validated experimentally and numerically. We use their analytical formula to predict the jumping performance of the popper, first, the maximum jumping height. 
The shell is initially in contact with the substrate via a ring and is lifted smoothly up to the transition. After the contact geometry transition, upon snapping the shell, the stored elastic energy of the flattened shell is converted into the kinetic energy of the popper. We presume that the maximum jumping height $H$ consists of two contributions:
$H_{\ell}$, representing the displacement of the shell as it recovers up to the transition, and $H_{\rm s}$, the maximum height of the vertical projectile driven by snap-buckling (Fig.~\ref{fig:2}(a-i)). 

The lifting height, $H_{\ell}$, corresponds to how high the shell is indented at the beginning of each experiment. In other words, the lifting height $H_{\ell}$ is equal to the distance between the height of the indented apex and that of the disk contact surface at the transition.
When the shell is fully inverted via mirror buckling~\cite{Pogorelov1988}, the apex displacement is geometrically predicted as $\simeq R(1 - \cos{\varphi})$, while the apex displacement at the transition is $e^* = \alpha_e h$. 
Combining these two contributions, the lifting height, $H_{\ell}$, is predicted as
\begin{eqnarray}
    H_{\ell} = R(1 - \cos{\varphi}) - \alpha_e h,\label{eq:Hl}
\end{eqnarray}
in excellent agreement with our experimental and numerical results, particularly for the range when the shell is largely indented as $\delta p_j/p_c \gtrsim0.3$ with the critical buckling-pressure, $p_c$,~\cite{Zoelly1915} (Fig.~\ref{fig:3}(a)). 

Right after the contact state changes from ring to disk-like geometry, the shell snaps and leaves the substrate by the height $H_{\rm s}$.
The travel distance obtained through snap-buckling, $H_{\rm s}$, is estimated by energy conversion. 
The critical state at the contact transition is realized when the elastic bending and stretching energies are balanced as $\mathcal{E} = F^*e^* (\sim Eh^3r^{*2}/R^2)$.
The stored elastic energy with the critical disk contact geometry, $\mathcal{E} = F^*e^*$, would be converted into the gravitational potential energy at the maximum height; $\mathcal{E} = mgH_{\rm s}$, from which we get
\begin{eqnarray}
    H_{\rm s} = \alpha_H \frac{Eh^4}{mgR}\label{eq:Hs}
\end{eqnarray}
with the prefactor, $\alpha_H= \alpha_F\alpha_e$, determined analytically.
Our analytical prediction correctly predicts experiments and simulations over an order of magnitude (Fig.~\ref{fig:3}(b)) without any fitting parameters.

The route of the maximum height is summarized as follows. 
The indented shell is lifted smoothly by the height $H_{\ell}$ with a decreasing ring contact radius up to the contact transition. The shell snaps when the distance between the shell apex and the substrate is of the order of the shell thickness as $w\sim h$, contributing to $H_{\rm s}$. Combining the formulas for $H_{\ell}$ and $H_{\rm s}$, we can fully predict the maximum jumping height as $H = H_{\ell} + H_{\rm s}$ without any fitting parameters, useful in predicting or even optimizing the jumping performance of the popper.

\subsection*{Snap \textit{or} Jump: Jumping condition of the popper}

{
We now study the jumping condition of the popper, i.e. whether the popper can jump or not. The outcomes of the jumping experiments are classified into three categories as summarized on the $(h/R)$-$(Eh^2/mg)$ plane in Fig.~\ref{fig:3}(c). \textit{Jump:} When the shell is thicker and stiffer, the net weight, $mg$, is much smaller than the impulsive force, $F^*$. In this case, the contact transition can offer a sufficient power source for the jump, whereas the launch of the popper fails for $mg>F^*$. \textit{No snap:} When the net weight is too large, the stored elastic energy of the fully-inverted shell $\sim Eh^{5/2}R^{1/2}$~\cite{Audoly2010} is insufficient to lift the popper up to $H_{\ell}$. Hence, the shell cannot snap at all, where the rim of the buckled shell remains in contact with the substrate even if $\delta p\simeq 0$. In between \textit{Jump} and \textit{No snap}, we find that the shell snaps (contact transition occurs), while it cannot leave the substrate, from which we categorize as \textit{Snap} state. Our theoretical predictions are consistent with both simulation and experiment, indicating that the stability conditions of the shell determine the jumping performance of the popper.}


We can conclude that the jumping process of the elastic shells is fully predictable. Throughout, we have found excellent agreement between experiments and simulations, validating the precision of the MPM simulations. Next, we clarify the role of friction between the shell and substrate, which is difficult to control systematically in experiments, fully relying on the simulation framework.

\subsection*{Roles of frictional properties of the substrate}

{
We find that the maximum jumping height decreases with the friction coefficient $\mu$, because friction suppresses the sliding of the contact surface~\cite{Sahli2024}. Given that the dimensionless friction coefficient is irrelevant to the physical scaling relation for the characteristic quantities, we expect that the friction coefficient, $\mu$, appears through the prefactors $\alpha_r, \alpha_F$, and $\alpha_e$. As shown in Fig.~\ref{fig:4}, all the prefactors characterizing the critical geometry at the transition decrease with $\mu$. The contact friction prevents the ring-to-disk transition, decreasing the critical size of the contact geometry. 
}

The prefactor of the snap-interval, $\alpha_t = \sqrt{3\alpha_e/\alpha_F}$ is also plotted in Fig.~\ref{fig:4}. Interestingly, $\alpha_t$ is almost constant against $\mu$, implying that the contact time scale for the jump is insensitive to the friction. 
The $\mu$-dependence of the prefactors derived here is applied to our other analytical formulas obtained in this paper, enabling us to predict the jumping phenomena of the popper on frictional substrates. 

\subsection*{Designing soft robots with predictive performance}


{
We have investigated the jumping dynamics of spherical shells (poppers) on a rigid substrate by combining experiments, simulation, and theory. Our predictive framework is based on the contact mechanics of shells with geometrically nonlinear deformation, allowing us to estimate their jumping performance without adjustable parameters. We have revealed that the transition of the contact geometry plays a critical role in the jumping dynamics. Our findings here would facilitate future studies for the dynamical performance of soft robots utilizing large deformation of slender structures with different geometry or energy inputs.}

{
Our study would be applied to designing novel soft robots with predictive performance. The building blocks of some climbing, hopping, or planetary robots, useful for working in unknown environments, consist of flexible components that undergo large deformation utilizing the stored elastic energy. Unraveling its mechanism is crucial both in predicting and optimizing the soft robots themselves. For example, the catapult of elastic shell investigated in this paper can be regarded as a geometric counterpart of latch-mediated spring actuation (LaMSA), where extremely fast motions and power amplification are realized in springy materials~\cite{Wang2023, Patek2023}. Our approach could be useful in predicting the performance of LaMSA-like structures leveraging flexible materials.  
}


%
There are several unsolved questions toward more predictive modeling of the jumping phenomena. 
We have analyzed the characteristic quantities (\textit{e.g.,} $r^*$) using the dimensionless prefactors with the upper and lower bounds. Their precise number would be computed once the dynamics and friction are considered in the previous analytical model~\cite{Audoly2010}, which we leave as an intriguing future theoretical work. 
Despite the nonlinearity of the experimental system, our simulation correctly predicts the experiments quantitatively, yet needs to be developed for more precise prediction. Simulating the fluid-structure interaction (\textit{e.g.,} flow fields in the container) will be a possible future extension, by which we expect to improve the time evolution of the container pressure or the launch time of the popper observed as the discrepancy from experiments.
We have studied the jumping performance of the popper on a flat frictional substrate. However, actual soft robots work in more complex environments, such as rough, inclined, or deformable surfaces. Their roles need to be investigated for further realistic predictions of soft robots in the future. 



\begin{figure}[h!]
    \centering
    \includegraphics[width =0.45\textwidth]{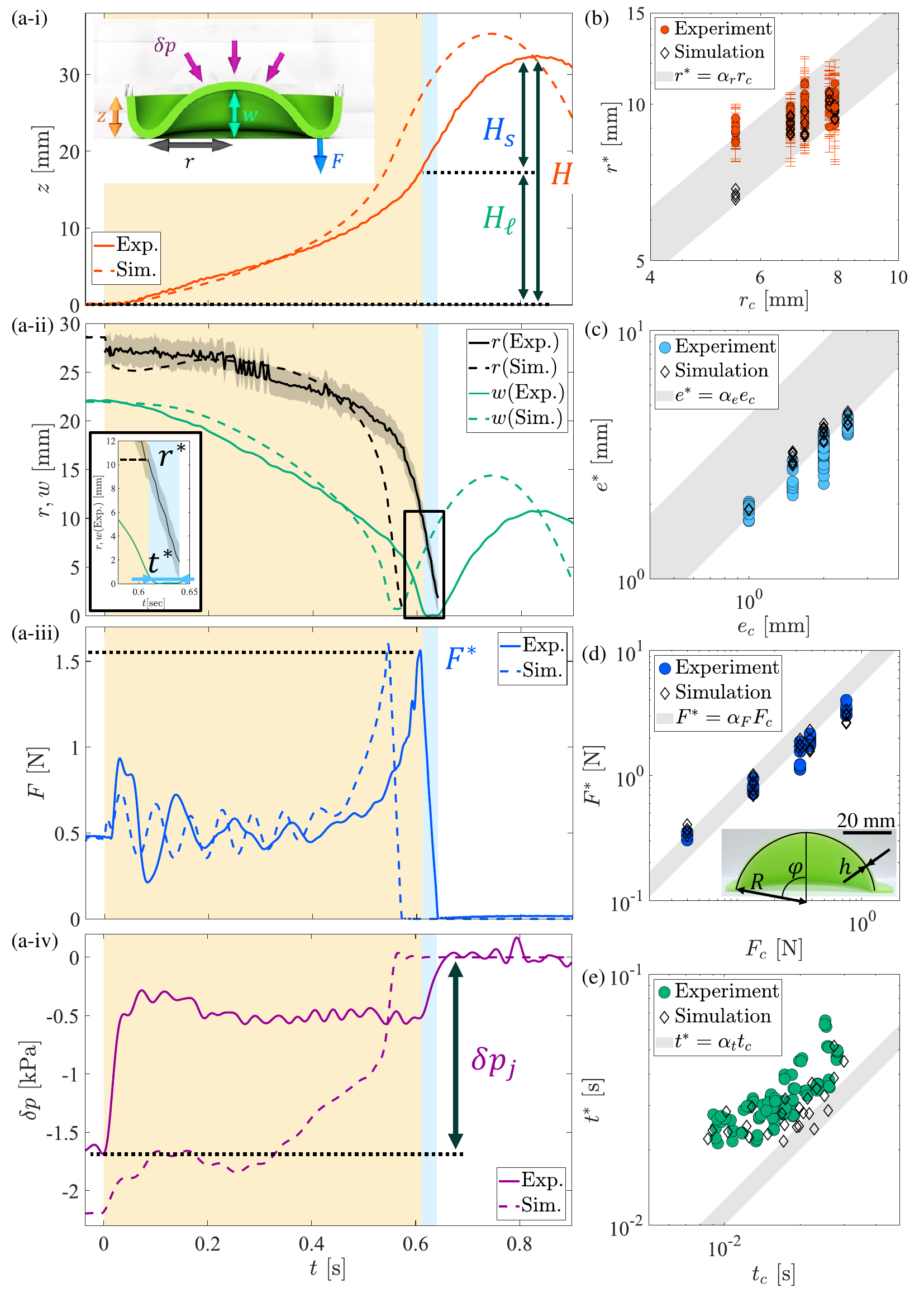}
    \caption{\textbf{Dynamical properties for the jumping process of shells.}(a-i)-(a-iv)~Typical time series data for the jumping process ($(R, h)= (30, 2.0)$~mm) in (solid lines) experiments and simulations (dashed lines)
    : (a-i) the vertical position of the container, $z$, (a-ii) contact radius (black), $r$ and the distance between the apex of the shell and substrate (green), $w$, (a-iii) reaction force of the substrate $F$, and (a-iv) the pressure difference between inner and atmospheric pressures, $\delta p$. The shaded error bars of $r$ represent the measurement error in the image analysis.
    The inset of (a-ii) is the zoomed experimental data for $r$ and $w$ near the contact transition. The experimental ring and disk contact states are highlighted by the orange and blue shaded regions, respectively.
    {(b)-(e) Characteristic quantities of the time series data are compared with the theory. Filled and empty data points represent the experiments and simulations, respectively.}
    (b) The radius of contact between shell and the substrate, $r^*$, and (c) the apex displacement, $e^*$, at the contact transition onset, (d) the maximum reaction force, $F^*$, and (e) the interval of disk-like contact, $t^*$ are plotted against the corresponding scaling predictions $r_c =\sqrt{hR}$, $e_c = h$, $F_c = Eh^3/R$, and $t_c = \sqrt{mR/Eh^2}$. The gray shaded regions represent our predictions as $r^* = \alpha_r r_c$, $e^* = \alpha_e e_c$, $F^* = \alpha_F F_c$, and $t^* = \alpha_t t_c$. The upper and lower boundaries of the shaded regions correspond to the theoretical upper and lower bounds of the prefactors predicted by analytical theory, respectively.
    }
    \label{fig:2}
\end{figure}

\begin{figure}[h]
    \centering
    \includegraphics[width =0.6\textwidth]{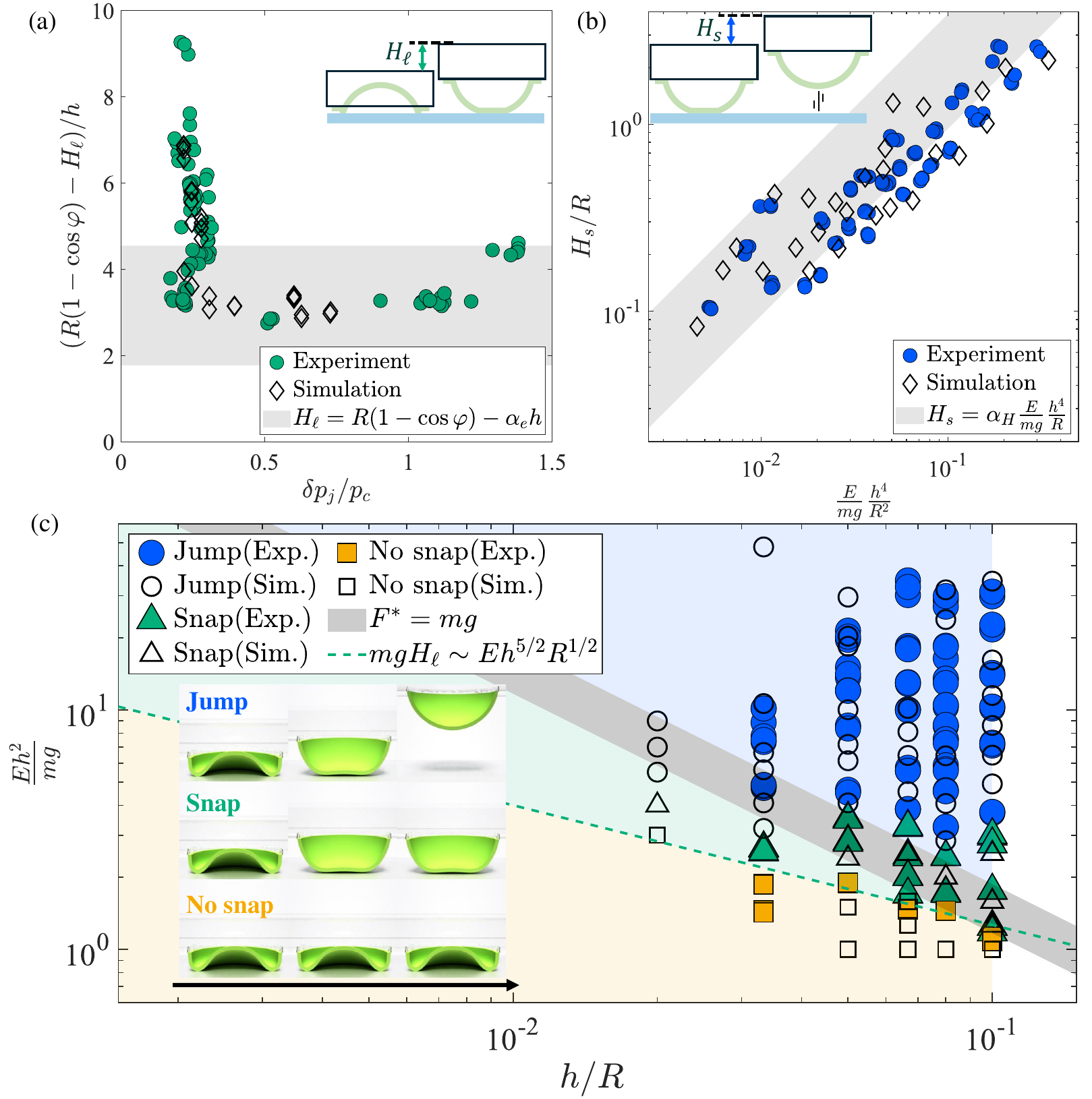}
    \caption{\textbf{Jumping performance of elastic shells.} (a)~Comparison of analytical formula Eq.(\ref{eq:Hl}) for $H_{\ell}$ with the experimental and numerical results. The horizontal axis represents the normalized pressure difference at the jump onset (defined as Fig.~\ref{fig:2}(a-iv)), $\delta p_j/p_c$. The shaded region represents the theoretical prediction in the fully inverted limit (large $\delta p_j/p_c$).
    (b)~Normalized height due to snap-buckling, $H_{\rm s}/R$, plotted against the scaling prediction, $Eh^4/(mgR^2)$. The shaded region represents the analytical prediction Eq.(\ref{eq:Hs}).
    (c)~Phase diagram of the jumping performance of shells, summarized on $(Eh^2/mg)$--$(h/R)$ plane.
    Circles, triangles, and squares represent the Jump, Snap, and No snap states, respectively (See insets for the schematics). The gray shaded and dashed lines represent our theoretical predictions.
    The corresponding filled and empty data points represent experimental and simulation results, respectively.
    }
    \label{fig:3}
\end{figure}

\begin{figure}[h]
    \centering
    \includegraphics[width =0.6\textwidth]{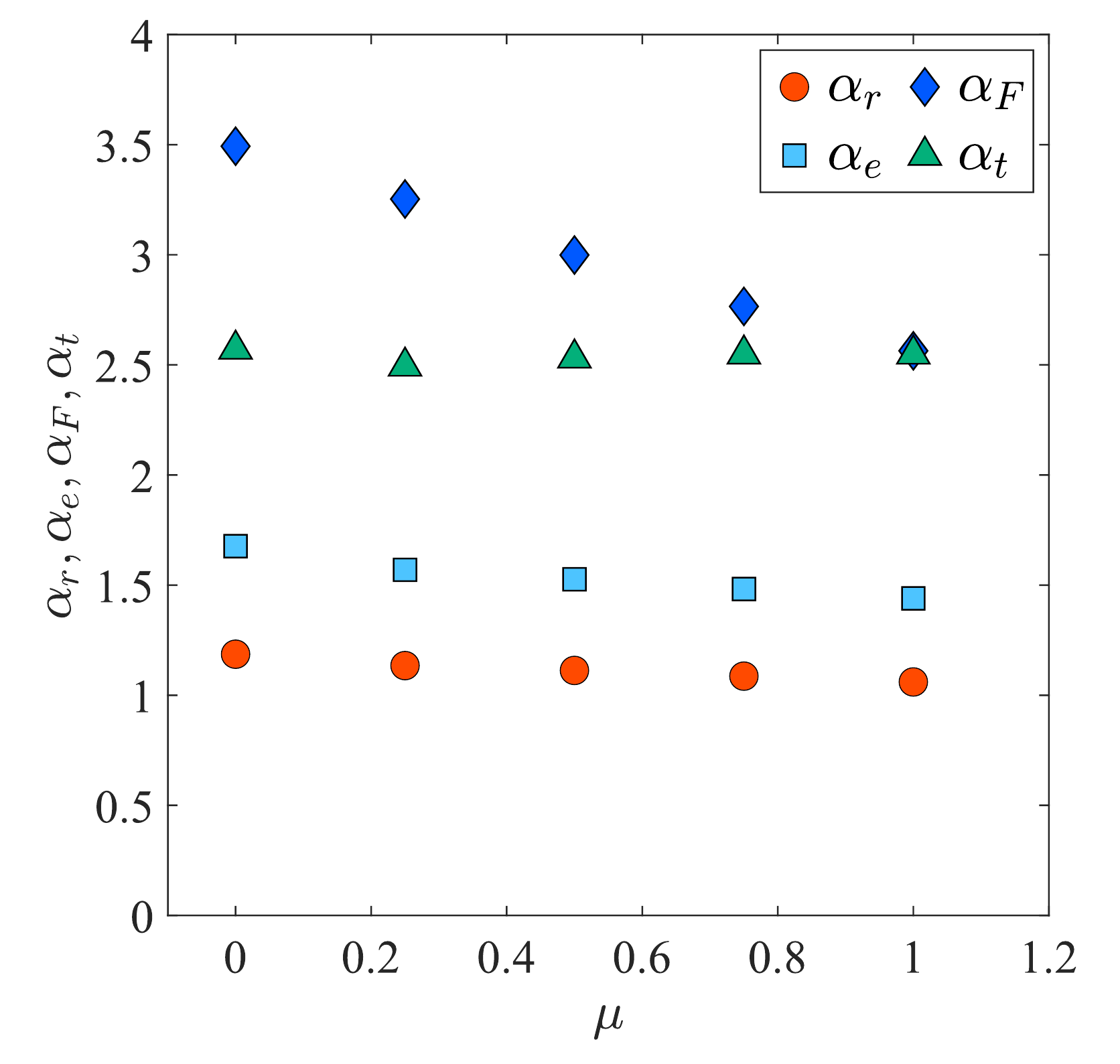}
    \caption{\textbf{The effects of the friction coefficient, $\mu$.} {The dimensionless prefactors, $\alpha_r, \alpha_e, \alpha_F$, and $\alpha_t$ obtained in simulations are plotted as a function of the friction coefficient between the shell and the substrate, $\mu$, for $(R,h) = (25,2.5)$mm.}}
    \label{fig:4}
\end{figure}


\clearpage 

%
\bibliography{science_template} 
\bibliographystyle{sciencemag}

%
%
%
%
%
%



\section*{Acknowledgments}
We thank Shuhei Shimizu for his discussion in the initial stage of this research. 
\paragraph*{Funding:}
This work was supported by MEXT KAKENHI 24H00299 (T.G.S.), 23K17317 (R.T.), JST FOREST Program, Grant Number JPMJFR212W (T.G.S.), JST the establishment of university fellowships toward the creation of science technology innovation Grant Number JPMJFS2125 (I.H.). 

\paragraph*{Author contributions:}
T.A., I.H., Y.N., S.K., R.T., H.T., G.I., and T.G.S designed the research and interpreted the results. T.A. performed experiments and analyzed the data, and T.G.S. supervised the experiments and theory. I.H. implemented the material point method, and S.K. and R.T. supervised the numerical investigations. Y.N. and H.T. designed and calibrated the force sensor. T.A., G.I., and T.G.S. designed and constructed the jumping experimental system.
T.G.S. managed the project. T.A., I.H., Y.N., S.K., R.T., H.T., G.I., and T.G.S. wrote the paper.

\paragraph*{Competing interests:}
There are no competing interests to declare.

\paragraph*{Data and materials availability:}
The data of all of the relevant plots of this paper have been compiled into a single Excel file and stored at the repository \url{https://doi.org/10.5281/zenodo.14264443}. This Excel file comprises 11 sheets, each corresponding to the data in Figures 2(a), 2(b)-(e), 3(a-c), 4, and S2(a,b).


\subsection*{Supplementary materials}
Materials and Methods\\
Supplementary Text\\
Figs. S1 to S3\\
References \textit{(50, \arabic{enumiv})}\\ 
Movie S1 to S3\\


\newpage


\renewcommand{\thefigure}{S\arabic{figure}}
\renewcommand{\thetable}{S\arabic{table}}
\renewcommand{\theequation}{S\arabic{equation}}
\renewcommand{\thepage}{S\arabic{page}}
\setcounter{figure}{0}
\setcounter{table}{0}
\setcounter{equation}{0}
\setcounter{page}{1} 


\begin{center}
\section*{Supplementary Materials for\\ \scititle}

	Takara Abe,
	Isamu Hashiguchi, 
	Yukitake Nakahara,
    Shunsuke Kobayashi, \\
    Ryuichi Tarumi, 
    Hidetoshi Takahashi,
    Genya Ishigami,
    Tomohiko G. Sano$^{*}$\\
\small{$^*${Corresponding author. Email: sano@mech.keio.ac.jp}}
\end{center}

\subsubsection*{This PDF file includes:}
Materials and Methods\\
Supplementary Text\\
Figures S1 to S3\\
Captions for Movies S1 to S3\\

\subsubsection*{Other Supplementary Materials for this manuscript:}
Movies S1 to S3\\

\newpage


\subsection*{Materials and Methods}

\subsubsection*{Fabrication of pneumatically-controlled elastic shells}
Spherical elastic shells are made of silicone elastomer (polyvinyl siloxane, Young's modulus $E$ = 1.2 MPa, Elite Double 32, Zhermack, Italy) and cast with 3D-printed molds. The base and catalyst solution of the liquid elastomer is mixed in equal weight using a centrifugal mixer (AR-100, Thinky Corporation, Japan) for 40~s at 2000~rpm (mixing) and another 60~s at 2200 rpm (degassing). The mixed solution is injected into the mold coated with the release agent (Ease Release 205, Smooth-On, USA) and degassed in the vacuum chamber to remove air bubbles. The shell is demolded after being cured (30 minutes). As a result, shells of five different aspect ratios are fabricated: ($R$, $h$) = (25, 2.5), (25, 2.0), (30, 2.0), (30, 1.5), and (30, 1.0) mm. The surface of the shell is treated with talcum powder (Baby Powder, Johnson \& Johnson, USA) to realize the dry frictional contact with the substrate. The surface-treated shell is glued with an acrylic container with a hole of $2R\sin{\varphi}$ diameter. The acrylic container is connected with a syringe (120~ml) via a flexible tube, allowing us to control the inner pressure of the container, $p$.

\subsubsection*{Measurement system}
The vertical location, $z$, contact radius, $r$, the distance between the shell apex and substrate, $w$, the reaction force of the substrate $F$, and the inner pressure of the container, $p$, are recorded simultaneously, and synchronized. We measure the vertical location of the container, $z$, and the distance between the apex and substrate, $w$, with two laser displacement sensors (IL-300, KEYENCE, Japan, 3~kHz sampling rate). One of the laser displacement sensors is set above the container (380~mm height from the rest height of the container) to measure $z$. The apex displacement of the shell is tracked by another sensor set below the force plate (described below). The laser passes the half mirror and the tiny hole (3~mm) at the center of the force plate. We gauge the inner pressure of the container, $p$, with the pressure sensor (AP-C30, KEYENCE, Japan, 400~Hz sampling rate). We build the custom-made force platform to measure the reaction force from the substrate precisely, similar to the methodology in \cite{Full1990}. Our force plate is made of a transparent square acrylic plate of 1~mm thickness (to visualize the shell contact) with a 3~mm diameter hole at the center (to measure $w$) surrounded by a square aluminum frame of 2~mm thickness. The frame is supported by four metallic legs with semiconductor strain gauges (KSPB-2-1K-E4, KYOWA, Japan), allowing us to measure the vertical force with a 12~mN resolution. The position error on the plate is within 2.6\%, the sampling rate is 5~kHz, and the resonant frequency is 170~Hz. 
The contact state between the shell and the force plate is visualized via a shadow illuminated by two high-power lights (LIUCWHA, Thorlabs, USA) and is recorded from below the force plate (through the half-mirror) by a high-speed camera (Phantom MIRO-C211, NOBBY TECH, Japan, 300~fps). The contact radius, $r$, is identified by performing edge detection of MATLAB against the recorded gray-scale images. The blurry edge of the shadow would cause possible measurement errors. We set the experimental measurement error of $r$ as the width of the initial ring contact highlighted by the shaded error bars in Fig.~\ref{fig:2}(a-ii) and the error bars of Fig.~\ref{fig:2}(b).

When the distance between the apex of the shell and the substrate, $w$, becomes sufficiently small relative to $R$ as $w/R<{0.03}$, we define that the contact transition (from ring-like to disk-like) starts.
The time when no force is applied to the substrate as $F\simeq0$ corresponds to the launch time of the shell, from which the interval of the areal contact $t^*$ is obtained.

\subsubsection*{Material Point Method}
We employ the material point method (MPM) to simulate the jumping process~\cite{Sulsky1995, Jiang2015, DEVAUCORBEIL2020185, Qin2023}. The elastic shell is represented as the set of a finite number of particles, while the force and moment balances are calculated on the background Eulerian grid. This hybrid numerical scheme enables us to simulate the large deformation with contact. We follow the identical protocol as experiments, adopting the experimental parameters, such as the shell geometry, mechanical property, and counterweight, except the proportional coefficient for the rapid airflow against the pressure difference of the container. 
The simulations are performed using the MPM implemented in Taichi{~\cite{Hu2019}}, which is a programming language that supports high-performance computing. 
See Supplementary Text more in detail.


\subsection*{Supplementary Text}




\subsubsection*{Details of Material Point Methods}

\begin{figure}[!h]
    \centering
    \includegraphics[width=0.6\linewidth]{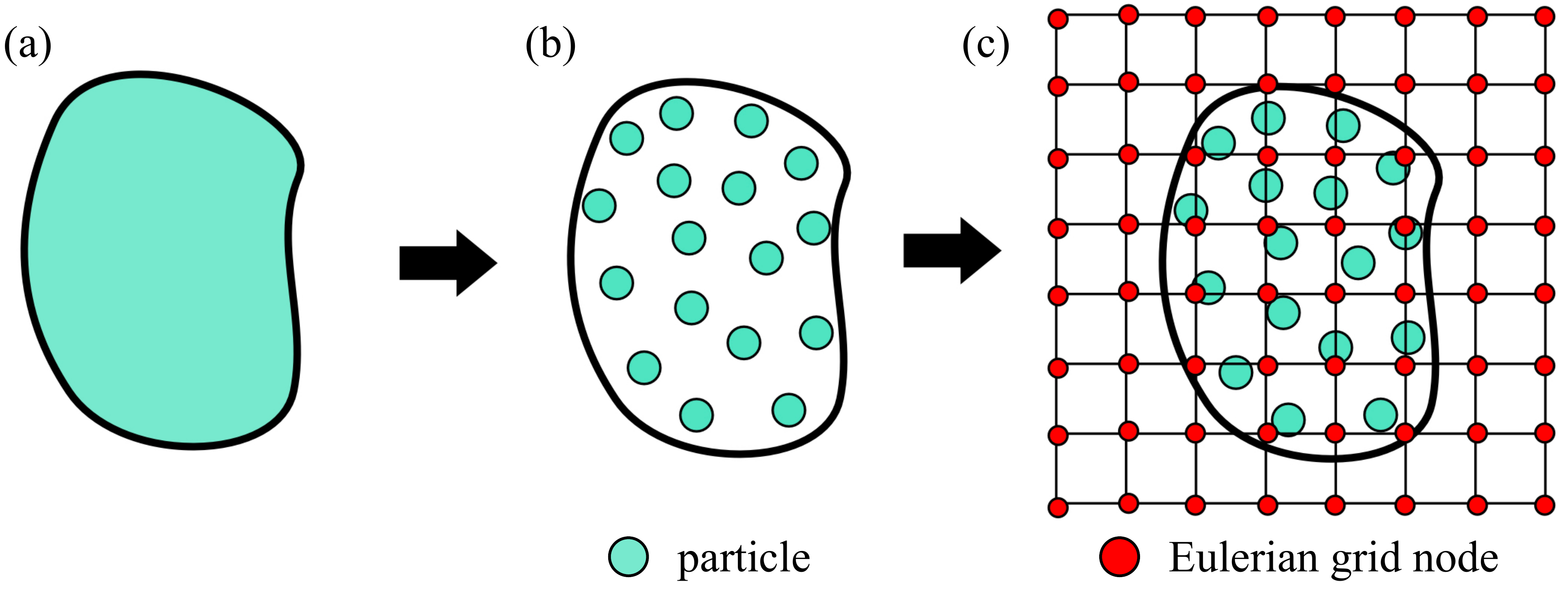}
    \caption{\textbf{Schematics of the MPM simulations.}(a)~The finite continuum is discretized into (b)~the set of particles. (c)~The physical quantities defined on the particle are distributed on the background Eulerian grid node. The internal and external forces and contact problems are calculated on the grid. We update the location of the particle based on the updated momentum. }
    \label{supp_fig:MPM}
\end{figure}

This section provides the mathematical details of the material point method, the hybrid computational method that simultaneously uses both material and spatial descriptions during the numerical integration of the equation of motion~\cite{Sulsky1995, Jiang2015}. The mathematical foundation is based on the Total Lagrangian Finite Element Method (TLFEM)~\cite{DEVAUCORBEIL2020185}.
Here, the time evolution of large deformations in a continuous medium is defined on the undeformed initial configuration, and the governing equation is derived using the calculus of variations. For discretizing the equation, we use standard finite element nodes for solids in addition to Eulerian grids in the computational domain.
Specifically, our elastic shell is represented as the set of a finite number of particles and is discretized onto the background Eulerian grids by introducing the basis functions. The equation of motion is integrated in time on the Eulerian grids, allowing us to take into account the complex contact boundary conditions in a straightforward manner (Fig.~\ref{supp_fig:MPM}).
This hybrid numerical scheme enables us to simulate the complex motions of soft robots, which undergo highly nonlinear large deformations with mechanical instability, contact, and friction~\cite{Qin2023}.

\textit{General framework:}~
We consider a simply connected compact set $\Omega_0$ in the standard Euclidean space $\mathbb{R}^3$.
The domain $\Omega_0$ is called the reference state describing the undeformed initial configuration.
The continuum is subjected to a non-infinitesimal displacement $\bm{u}=\bm{u}(\bm{x},t)$, which is differentiable with respect to its arguments as many times as required.
Let $\dot{\bm{u}}=\bm{v}$ be the material time derivative of the displacement.
Then, the Lagrangian functional $\mathcal{L}$ for the continuum is expressed as follows:
\begin{align}
\label{lagrangean function}
\mathcal{L}\left(\bm{u},\dot{\bm{u}}\right) = T\left(\dot{\bm{u}}\right) - W\left(\bm{u}\right) + \Pi\left(\bm{u}\right).
\end{align}
The functions $T$, $W$, and $\Pi$ represent the kinetic energy, strain energy, and potential energy due to external forces integrated over the time domain $[t_0, t_1]$ defined as
\begin{align}
\label{Kinematic_Energy_Function}
T(\dot{\bm{u}}) &\equiv \int_{t_0}^{t_1} \int_{\Omega_0} \frac{1}{2} \rho_R \left( \dot{\bm{u}} \right)^2 dV dt,
\\
\label{Stranin_Energy_Function}
W(\bm{u}) &\equiv \int_{t_0}^{t_1} \int_{\Omega_0}\left(\frac{\mu}{2} (\text{tr}\left(\textbf{F}^T\textbf{F}\right) - 3) - \mu \mathrm{log}(J) + \frac{\lambda}{2} \mathrm{log}^2(J)\right) dV dt,
\\
\label{Force_Energy_Function}
\Pi(\bm{u}) 
&\equiv\int_{t_0}^{t_1} \int_{\Omega_0} \bm{b}\cdot\bm{u}\,dV dt + \int_{t_0}^{t_1} \int_{\Gamma_{N} }\bar{\bm{\tau}}\cdot\bm{u}\,d\Gamma dt.
\end{align}
Here, $\rho_R$ is the mass density in the reference state and $\lambda$ and $\mu$ are called the Lam\'e constants.
$\textbf{F}$ is the deformation gradient tensor, and $J=\text{det}(\textbf{F})$ is the volume change ratio between the reference and current states.
The strain energy $W$ is defined using the Neo-Hookean hyperelastic material in three dimensions, which describes the deformation of elastomers.
The potential energy $\Pi$ of external forces is defined using the body force $\bm{b}$ acting directly on a small element $dV$ in the reference state and the surface force $\bar{\bm{\tau}}$ applied to the Neumann boundary, \textit{e.g.}, the pressure difference of the popper.

The governing equation in the weak form can be derived under the condition that the first variation of the Lagrangian functional is zero $\mathcal{L} = 0$ as: 
\begin{align}
\label{weak momentum equation} \int_{t_1}^{t_2}\int_{\Omega_0}\rho_R\ddot{\bm{u}}\cdot\delta\bm{u} dVdt
= &
-\int_{t_1}^{t_2}\int_{\Omega_0}\left(\mu (\textbf{F} - \textbf{F}^{-T}) + \lambda \mathrm{log}(J)\textbf{F}^{-T}\right):\delta \textbf{F} dVdt \nonumber
\\
& + \int_{t_1}^{t_2} \int_{\Omega_0} \bm{b}\cdot\delta \bm{u} dV dt + \int_{t_1}^{t_2} \int_{\Gamma_N} \bar{\bm{\tau}} \cdot\delta\bm{u} d\Gamma dt
\end{align}
The weak form of the equation of motion (\ref{weak momentum equation}) defined for the continuum is discretized into that for the particles (Lagrange representation), which is further rewritten into the form for the Eulerian grid utilizing the basis (interpolation) functions. For example, the displacement vector $\bm{u}$ and its variation $\delta\bm{u}$ are approximated as
\begin{align}
\label{approx u}
\bm{u}(\bm{x},t) &= \sum_{i} K_{i}(\bm{x})\bm{u}_{i}(t),
\\
\label{approx delta u}
\delta \bm{u}(\bm{x},t) &= \sum_{i} K_{i}(\bm{x})\delta\bm{u}_{i}(t).
\end{align}
Note that $K_i(\bm{x})$, $\bm{u}_{i}$, and $\delta\bm{u}_{i}$ 
represent the finite element shape function of the quadratic tetrahedral element~\cite{wriggers2008}, the displacements and its variation of the $i$-th particle.

By setting the stationary condition, $\delta\mathcal{L}=0$, for any variation $\delta\bm{u}$,
the weak form of the equation of motion (\ref{weak momentum equation}) can be rewritten into the following discretized equations for {$i$-th particle}:
\begin{align}
{m_i\ddot{\bm{u}_i} = \bm{f}_i^{\text{ext}} + \bm{f}_i^{\text{int}}}\label{eq:eom_i}
\end{align}
where, the {lumped} mass, ${m_i}$, the internal, $\bm{f}_{{i}}^{\mathrm{int}}$, and external forces, $\bm{f}_{{i}}^{\mathrm{ext}}$, are respectively defined as
\begin{align}
m_{i}
&\equiv
\frac{\int_{\Omega_0}K_i^2dV}{\int_{\Omega_0}\sum_j K_j^2dV}\int_{\Omega_0}\rho_R dV
\\
\label{eq:fint}
\bm{f}_{{i}}^{\mathrm{int}}
&\equiv
\int_{\Omega_0} -\left(\mu (\textbf{F} - \textbf{F}^{-T}) + \lambda \mathrm{log}(J)\textbf{F}^{-T}\right) \frac{\partial K_{{i}}}{\partial \bm{x}} dV,
\\
\label{eq:fext}
\bm{f}_{{i}}^{\mathrm{ext}}
&\equiv
\int_{\Omega_0} \bm{b} K_{{i}} dV + \int_{\Gamma_N} \bar{\bm{\tau}} K_{{i}} d\Gamma.
\end{align}
Note that we do not sum over the indices $i$ in the left-hand side of Eq.~(\ref{eq:eom_i}).
In this study, a dynamic explicit method is applied to the discretization in the time direction.

\textit{Coupled Analysis with Internal Fluid:}~
{From the equation of state for the isothermal process, the pressure difference of the popper from outside, $\delta p$,} is calculated as follows:
\begin{align}
\label{eq:P_inner}
{\delta p} = p_0\left(\frac{J_{\text{n}}}{J_{V}} - 1\right)
\end{align}
with the atmospheric pressure outside the popper, {$p_0$}. We introduce the volume and mass change ratio of the interior of the popper, $J_{\rm V}$ and $J_{\rm n}$, respectively, defined as
\begin{align}
    J_{\text{V}} = \frac{V_{\text{in}}}{V_{\text{in}}^0},
    \quad
    J_{\text{n}} = \frac{n_{\rm in}}{n_{\rm in}^0}.
\end{align}
Note that $V_{\text{in}}^0$, $V_{\text{in}}$, $n_{\rm in}^0$ and $n_{\rm in}$ represent, the initial and current volumes of the popper, and the initial and current mass of the air inside the popper, respectively. 
We adopt the empirical linear dynamics for the mass change inside the popper as $\ddot{J}_{\rm n} = -\alpha_{\text{n}}{\delta p}$, {and the initial value $\dot{J}_{\text{n}}(t=0)=-\beta_{\text{n}}\delta p(t=0)$}  with the fitting parameters $\alpha_{\text{n}}$ {and $\beta_{\text{n}}$}. The fitting parameters introduced here are characteristic of the experimental system and set $\alpha_{\text{n}} = 1.0\times 10^{-3}$ {[Pa${}^{-1}\cdot$s${}^{-2}$] and $\beta_{\text{n}} = 5.0\times 10^{-5}$ [Pa${}^{-1}\cdot$s${}^{-1}$]} throughout. The experimental measurement of $\alpha_{\text{n}}$ together with the validation of the pressure rule in Eq.~(\ref{eq:P_inner}) is beyond the scope of the current study, which we leave as a future work. 
The current pressure is updated against the 
volume $J_{\rm V}$ and mass change $J_{\rm n}$, following Eq.~(\ref{eq:P_inner}). 
Using the updated pressure, the deformation of the popper is calculated, and then the current volume of the popper is updated. In the discretized formulation Eq.~(\ref{eq:eom_i}), the pressures acting on the popper appear as the surface force $\bar{\bm{\tau}}$ in Eq.~(\ref{eq:fext}). 

\textit{Contact analysis by MPM:}~
The MPM method is a hybrid computational framework combining Lagrange and Euler representations of the continuum, whose advantage is that one can handle the complex contact mechanics with different continua or rigid bodies, critical in the jumping process of the popper studied in the main text. This advantage originates from the fact that the equation of motion is integrated on the Eulerian grid, while the artificial adhesive forces appear in the standard MPM method (due to the fact that the relative particle velocities near the contact surfaces are small). The effect of such numerical artifacts can be reduced through the following procedure. We compute the surface velocities of the continua. If the normal component of the relative velocity is negative (when two continua are approaching), the contact force is calculated. If not (if they are releasing), the contact force is set to be zero.

We detail the procedure to compute the contact force (when two continua are approaching). The procedure consists of the following three steps. 1:~The velocity and mass of the particles are distributed onto the neighboring nodes.
2:~The contact condition is examined on the grid.
If the continua and rigid plate are approaching, the contact forces are calculated in this step. 3:~The contact forces calculated on the node are mapped onto the particles. More detailed procedures follow below.

\begin{enumerate}
    \item \textbf{Particle to the Euler grid mapping:} In the mapping process to the Eulerian grid, the velocity, $\bm{v}_i$, and mass, $m_i$, on the $i$-th particle (nodal point) are distributed to the neighboring $\alpha$-th Eulerian grid node via the basis function, $N_{\alpha}(\bm{x})$, as follows. Let $n$ be the number of the time step. We write the quantities with the superscript $n$ as those at time step $n (=1,2,\cdots)$. 
    Based on the particle configuration in the previous ($n-1$-th) time step, $\bm{x}_i ^{n-1} (i=1,2,\cdots)$, we calculate the internal and external forces (Eqs.~(\ref{eq:fint}) and (\ref{eq:fext})) and calculate the current ($n$-th) momentum of the $i$-th particle, $\bm{p}_i ^n$ by integrating the equation of motion Eq.~(\ref{eq:eom_i}).
    Using the current the momentum, $\bm{p}_i ^n$, and mass on the $\alpha$-th Eulerian node, $m_{\alpha}^{n}$, are approximated as
    \begin{align}
        \label{p2g mapping momentum}
        \bm{p}_{\alpha}^{n} &= \sum_i m_i\left(\bm{v}^{\prime}_i + \textbf{C}^{n-1}_i(\bm{x}_{\alpha} - \bm{x}_i ^{n-1})\right)N_{\alpha}(\bm{x}_i ^{n-1}-\bm{x}_{\alpha}),\\
        \label{p2g mapping mass}
        m_{\alpha}^{n} &= \sum_i {m}_iN_{\alpha}(\bm{x}_i ^{n-1}-\bm{x}_{\alpha}),
    \end{align}
    where the summation $\sum_i$ runs over all the particles, and $\bm{v}_i^\prime$ represents the velocity updated from $\bm{v}_i^{n-1}$ using Eq. (\ref{eq:eom_i}).
    Note that the variables with the subscript $i$ or $\alpha$ are those defined on the particle and Euler grid, respectively. For example, $\bm{x}_i$ and $\bm{x}_{\alpha}$ represent the locations of $i$-th particle and $\alpha$-th grid node.
    In Eq.~(\ref{p2g mapping momentum}) we adopt the Affine Particle-in-cell (APIC)\cite{Jiang2015} calculation scheme, which is characterized by the affine velocity tensor, $\textbf{C}_i^n$, the second rank tensor defined using the previous particle configuration (updated at the 3rd step of the contact calculation later).
    \item \textbf{Update of the momentum on the Eulerian grid:} The current momentum on the grid, $\bm{p}_{\alpha} ^n$, is updated temporarily as $\tilde{\bm{p}}_{\alpha} ^n$ by calculating the contact conditions and forces. We calculate the updated momentum of the grid as follows, depending on whether it satisfies the Dirichlet boundary condition, contact boundary condition, or neither of them.
    \begin{itemize}
        \item
        \textit{Dirichlet boundary condition:}
        {The momentum $\bm{p}_{\alpha}^n$ in the neighborhood of particles on the clamped edge of the shell is set to be identical to the momentum of the rigid body, as $\bm{p}_{\alpha}^n = v_z^{\text{rigid}}\bm{e}_z$}, where $v_z^{\text{rigid}}$ and $\bm{e}_z$ represent the velocity of the rigid body to which the shell is attached (container) and the basis vector in the direction of the jump. The velocity of the rigid body, $v_z^{\text{rigid}}$, follows the equation of motion, where all the external forces are taken into account (including the counterweight).
        \item 
        \textit{Contact boundary condition against {the rigid plate}:}
        If the contact condition {$\bm{p}^{n}_{\alpha}\cdot\bm{e}_z < 0$} is satisfied (two surfaces are approaching), the continuum boundary is judged to penetrate the {rigid plate}. We then update the momentum by subtracting the momentum in the direction of a jump as
        {
        \begin{align}
            \tilde{\bm{p}}_{\alpha}^n = \bm{p}^{n}_{\alpha} - (\bm{p}\cdot\bm{e}_z)\bm{e}_z,\label{eq:pndiff}
        \end{align}
        }
        The renewed momentum $\tilde{\bm{p}}_{\alpha}^n$ would be used to calculate the location of the particle, $\bm{x}_{i} ^{n+1}$, in the next time step. 
        Note that the dynamic friction force with the boundary is calculated by identifying the normal force, $F_{\rm norm}$, as the change of the momentum in the direction of jump {during the single time step} as {$F_{\rm norm} \equiv (\tilde{\bm{p}}_{\alpha}^n - \bm{p}^{n}_{\alpha})\cdot\bm{e}_z/\Delta t$, where $\Delta t$ is the time step size}. The magnitude of the dynamic friction force is then given by $\mu F_{\rm norm}$, which is taken into account to the update of the tangential component of the momentum with the kinetic friction coefficient, $\mu$.
        The momentum update by this process is only performed on the three-layer Euler nodes that satisfies
        $\bm{x}_{\alpha}\cdot\bm{e}_z = 0$, $-\Delta x$, $-2\Delta x$ with the spacing of the Eulerian grid, $\Delta x$.
        \item \textit{Neither under the Dirichlet or contact boundary conditions:} If the Euler grid is not subject to the Dirichlet or contact boundary conditions, the momentum is set to be $\tilde{\bm{p}}_\alpha^n=\bm{p}_\alpha^n$. Typically, these grids correspond to the free boundary or the internal of the continuum.
    \end{itemize}
    \item \textbf{Euler grid to Particle mapping:} 
    The updated momentum calculated on the Eulerian grid, $\tilde{\bm{p}}_{\alpha} ^n$, is calculated back to the particle. Using the current temporal momentum, $\tilde{\bm{p}}_{\alpha} ^n$, and the mass, $m_{\alpha} ^n$, we define the current velocity on the Eulerian grid, $\bm{v}_{\alpha} ^n$, as $\bm{v}_{\alpha} ^n \equiv \tilde{\bm{p}}_{\alpha} ^n/m_{\alpha} ^n$. We then map the Eulerian velocity onto the particle velocity, using the basis function, $N_{\alpha}(\bm{x})$, as
    \begin{align}
        \label{g2p velocity}
        \bm{v}_i^{n} &= \sum_{\alpha} \bm{v}_{\alpha} ^n N_{\alpha}(\bm{x}_{\alpha}-\bm{x}_i ^{n-1} ).
    \end{align}
    The location of the particle is integrated as $\bm{x}_i^{n} = \Delta t \bm{v}_i^{n}$.
    Using the current configurations, we calculate the affine velocity tensor for the next time step, $\textbf{C}_i^{n}$, as
    \begin{align}
        \textbf{C}_i^{n} &= \frac{4}{\Delta x^2}\{\bm{v}_{\alpha}^{n}\otimes\left(\bm{x}_{\alpha} - \bm{x}_i^{n-1}\right)\}N_{\alpha}(\bm{x}_{\alpha}-\bm{x}_i ^{n-1}),
    \end{align}
    The affine velocity tensor calculated here will be used in the step 1 of the next time step $n+1$.
\end{enumerate}

\subsubsection*{Elastic shells pushed by a rigid plane}
{In this section, we describe the procedure to determine the dimensionless coefficients $\alpha_F, \alpha_r$ and $\alpha_e$. We rely on the analytical prediction detailed in Ref.~\cite{Audoly2010}, where the force-displacement curve for the elastic shells pushed by a rigid plane is derived analytically within the shallow shell theory. It has been reported that there exists the hysteresis in the force-displacement curve, upon compression/decompression cyclic tests of the rigid plane, following the contact geometry transition from disk to ring and ring to disk, respectively. Our jumping experiment studied in the main text corresponds to the inverse protocol (\textit{decompression process}), where the contact geometry transits from the ring to the disk. We first review the previous theoretical results briefly~\cite{Audoly2010} together with its experimental validation and then describe the procedure to determine the dimensionless prefactors, $\alpha_F, \alpha_r$, and $\alpha_e$.}

{When the shell and the substrate are in contact via the disk geometry, the force, $F_{\rm disk}$, and the displacement of the shell apex, $e$, are related via
\begin{eqnarray}
    F_{\rm disk} = \frac{Eh^3}{12(1-\nu^2)R}\hat{N}(\hat{e})\label{eq:Fe_disk},
\end{eqnarray}
where the dimensionless function, $\hat{N}=\hat{N}(\hat{e})$, is a convex function of $\hat{e}$ and is computed numerically as in Ref.~\cite{Audoly2010}, with the rescaled apex-displacement $\hat{e} = \sqrt{12(1-\nu^2)}(e/h)$ and the Poisson ratio $\nu$. The force increases rapidly with $e$ because the disk contact with the large radius, $r (\simeq R)$, is unfavorable energetically. The contact geometry would transit to the ring geometry above the critical loading force. When the shell and the substrate are in the ring contact with the radius, $r = r(e)$, the force for the ring contact, $F_{\rm ring}$, is expressed as
\begin{eqnarray}
    F_{\rm ring} = \frac{Eh^{5/2}n_{(2)}}{\{12(1-\nu^2)\}^{3/4}R^{3/2}} r({e})\label{eq:Fe_ring},
\end{eqnarray}
where $n_{(2)}$ is numerically calculated as $n_{(2)} = 20.83$~\cite{Audoly2010}. From geometry, the contact radius $r$ and the apex displacement $e$ are related as $r = \sqrt{eR}$. Hence, we rewrite Eq.~(\ref{eq:Fe_ring}) as
\begin{eqnarray}
    F_{\rm ring} = \frac{Eh^{3}n_{(2)}}{\{12(1-\nu^2)\}^{3/4}R}\sqrt{\frac{e}{h}}.
\end{eqnarray}
Note that $F_{\rm ring}=F_{\rm ring}(e)$ is a concave function of the apex displacement, $e$, and that the ring contact with small $e$ ($e/h\lesssim 1$), $F_{\rm ring}\sim (e/h)^{1/2}$, requires a larger force than that for the disk contact Eq.~(\ref{eq:Fe_disk}).
}

{The linear stability analysis of the disk contact state reveals that the configuration with Eq.~(\ref{eq:Fe_disk}) becomes unstable when the force for the forward protocol (increasing $e$) reaches the critical value as $F_{\rm disk} = F_{+}$ ($\hat{N}(\hat{e})=N_+ (\simeq76.94)$), above which the contact shape would transit to the ring geometry. 
Decreasing the displacement $e$ from the ring contact configuration, the contact geometry transits from the ring to the disk (\textit{inverse protocol}). In other words, there is some hysteresis in the force-displacement curve, where both disk and ring contacts are stable for the range of the force below $F_{+}$ as $F_{-}<F<F_{+}$. The lower band of the hysteresis, $F_{-}$, could be computed numerically by constructing the full numerical solution~\cite{Audoly2010}. Nevertheless, we attempt to speculate $F_{-}$ by finding the apex displacement, $e>0$, satisfying $F_{\rm disk} = F_{\rm ring}$, which gives $F_{-} = {Eh^3}{N}_{-}/\{12(1-\nu^2)R\} $ with $N_- \equiv 46.8$ in Eq.~(\ref{eq:Fe_disk}). 
}

\begin{figure}[h]
    \centering
    \includegraphics[width =0.8\textwidth]{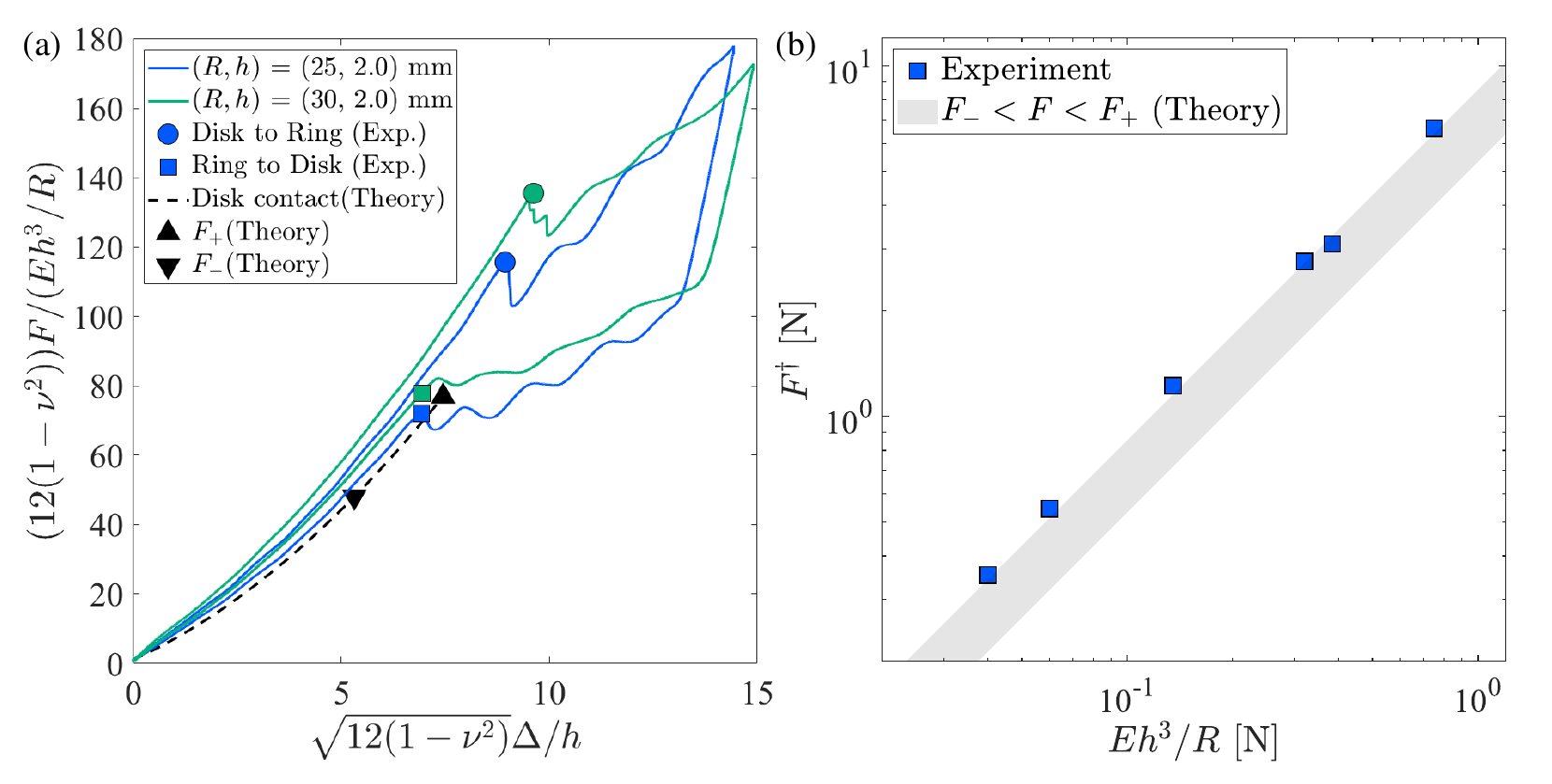}
    \caption{\textbf{Experimental results of elastic shells pushed by a rigid plane, compared with the analytical theory.}
    (a) The rescaled force-displacement curves for different aspect ratios $h/R$. The vertical and horizontal axis are normalized by the characteristic quantities. The colored circles and squares represent the experimental disk-to-ring and ring-to-disk transition points, respectively. The dashed line represents the analytical prediction in Ref.~\cite{Audoly2010} with the upper and lower bounds of the theoretical coexistence region. 
    (b) The experimental critical force for the ring-to-disk transition points, $F^{\dagger}$, (squares in (a)) superposed with the theoretical coexistence limit, $F_-<F<F^+$. 
    }
    \label{supp_fig:2}
\end{figure}

{
We perform cyclic compression/decompression tests of shells using the rigid plate for several aspect ratios $h/R$ to validate the above theoretical predictions. We observe the hysteretic force-displacement curves as predicted theoretically~\cite{Audoly2010} and reported experimentally~\cite{Pauchard1997} (Fig.~\ref{supp_fig:2}(a)) with the displacement of the plate, $\Delta$. Upon compression, the force increases along with Eq.~(\ref{eq:Fe_disk}) and transits to the branch with the ring contact, while the decompression curve follows the different lower branch from that for the compression and merges with the disk contact prediction at another transition point $F^{\dagger}$ (defined as the squares in Fig.~\ref{supp_fig:2}(a)). The theoretical upper and lower bounds for the coexistence of disk and ring contacts, $F_+$ and $F_-$, are superposed in the rescaled force-displacement curve (Fig.~\ref{supp_fig:2}(a)), which underestimate experimental values but are still close to the experimental ring-to-disk transition points, $F^{\dagger}$. It should be noted that there are no adjustable parameters for Eq.~(\ref{eq:Fe_disk}) plotted as the dashed line in Fig.~\ref{supp_fig:2}(a). We plot the experimental lower bounds, $F^{\dagger}$, for several $h/R$ as a function of $Eh^3/R$ and compare with the theoretical coexistence limit $F_{-}<F<F_{+}$ highlighted by the shaded region in Fig.~\ref{supp_fig:2}(b). Again, the theory still underestimates the experimental values, while the agreement is excellent at the scaling level.  
}

{
The possible mechanism for the discrepancy of the dimensionless prefactors would be, for example, the inevitable experimental defects (\textit{e.g.,} air bubbles in the curing of elastomer), contact friction, and the dynamical properties, which are not incorporated in theory. Taking into account the above contributions in the theoretical framework may improve the prediction for the pre-factors, while it is highly challenging. Hence, we leave further theoretical studies as one of the future works.
}

{
In the main text, to establish the fully quantitative predictions for the jumping height and the phase diagram, we introduce the dimensionless pre-factors (\textit{e.g.,} $\alpha_F$) whose theoretically acceptable minimum and maximum values are estimated via the theoretical lower $F_-$ and upper bounds $F_+$, respectively, giving the prediction as $F^* = \alpha_F F_c$ with $5.3<\alpha_F<8.5$ (shaded regions in Fig.2(d) of the main text), which is surprisingly excellent agreement with experiments and simulations. 
The possible ranges of other prefactors are determined through similar strategies. 
The critical contact radius $r^*$ and the apex displacement $e^*$ are estimated as $r^*= \alpha_r r_c$ and $e^* = \alpha_e h$ with
$1.2<\alpha_r<1.6$ and $1.8<\alpha_e<4.5$. The time interval for the disk contact, $t^* = \alpha_tt_c$, is estimated by integrating the equation of motion for the apex displacement, $e = e(t)$: $md^2e/dt^2 = F^*\{1- (t/t^*)\}$ for the interval $t^*$, leading to the relation as $\alpha_t = \sqrt{3\alpha_e/\alpha_F}$.
The estimated prefactors, $\alpha_F, \alpha_r, \alpha_e$, and $\alpha_t$, are utilized to estimate the further theoretical predictions for $H_{\ell}$ and $H_{\rm s}$ as detailed in the main text. 
}

\subsubsection*{Asymmetric buckling of shells}

\begin{figure}[h]
    \centering
    \includegraphics[width =0.7\textwidth]{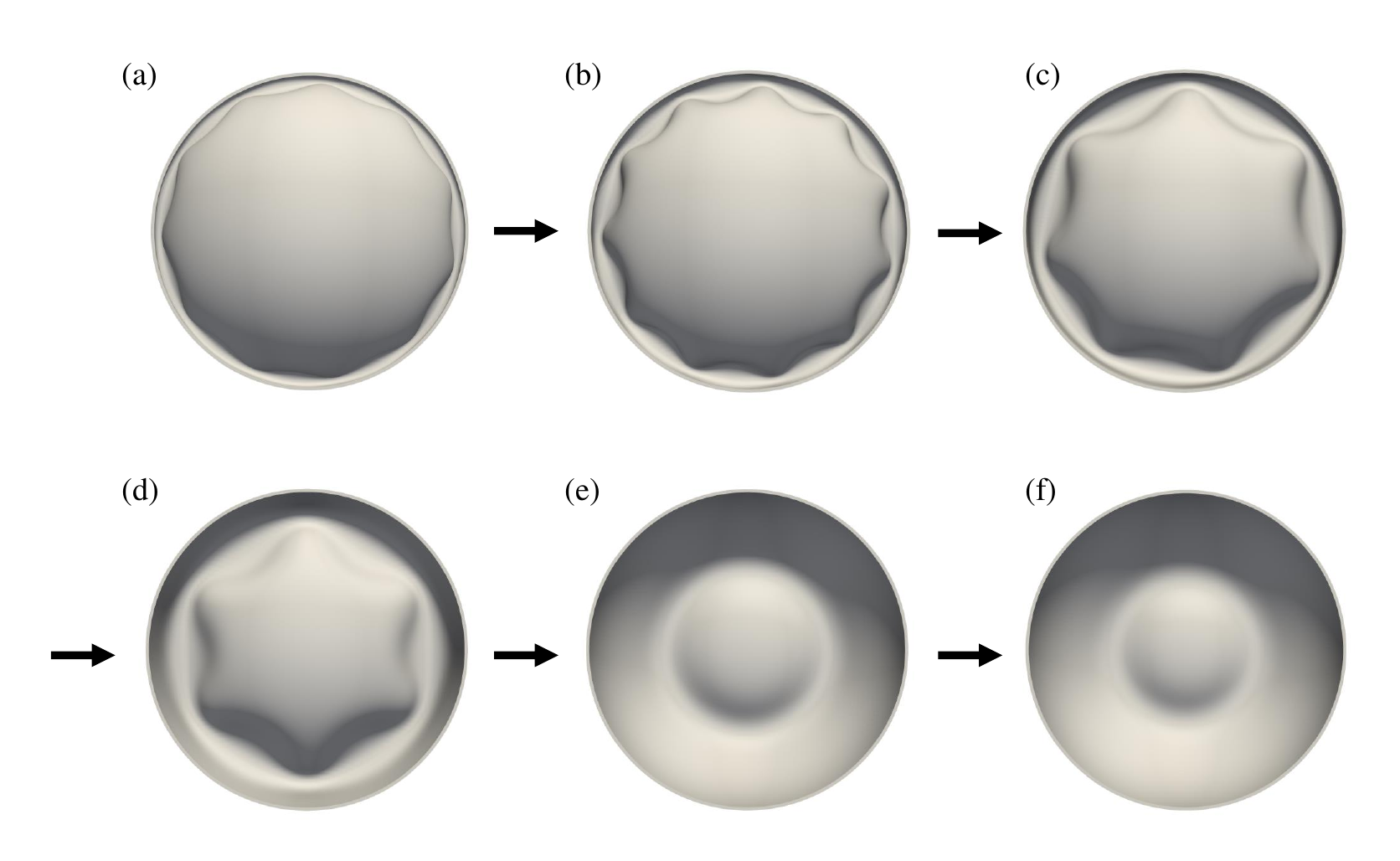}
    \caption{\textbf{Simulation results for the asymmetric buckling of shell}
    (a)-(f)~The top views of the shell that undergoes the asymmetric buckling obtained in the simulation with $h/R=0.02, Eh^2/mg = 3.0$. The number of vertices decreases as the shell recovers its original shape.
    }
    \label{supp_fig:3}
\end{figure}

Our theoretical framework is based on the axisymmetric snap-buckling of shells, while asymmetric buckling with multiple vertices is often observed in the buckling of thin shells~\cite{Nasto2013}. Indeed, we observe asymmetric buckling for thinner shells at the initialization of the shell before jumping (Fig.~\ref{supp_fig:3}(a)). However, the asymmetric buckling is irrelevant to the jumping process, because the multiple vertices disappear as the shell recovers its natural shape. We show the series of snapshots (top view) of the shell of $h/R=0.02$ in the No Snap state, where the shell is in the ring contact even if $\delta p\simeq 0$. The shell is asymmetrically buckled initially with a polygonal rim (Fig.~\ref{supp_fig:3}(a)), whereas the number of vertices decreases, and the rim finally becomes a cylindrical shape (Fig.~\ref{supp_fig:3}(e)). Hence, we believe that asymmetric buckling is not crucial in the jumping process.


\clearpage 

\paragraph{Caption for Movie S1.}
\textbf{Movie of the jumping moment in the experiment and simulation.}
(left)~The experimental movie of the jumping phenomena for the experimental shell with $(R, h) = (25, 2)$~mm. (right)~The corresponding simulation movie of the shell cross-section.

\paragraph{Caption for Movie S2}
\textbf{Experimental sequence and the corresponding time evolution of the container pressure.}
The animated version of the experimental sequence. 

\paragraph{Caption for Movie S3}
\textbf{Time evolution of the contact geometry and the mechanical performance of the popper.}
The time-synchronized data for the apex height, $w$, popper height, $z$, contact radius, $r$, contact force, $F$, the inner pressure difference, $\delta p$. The change of the contact geometry taken by the high-speed camera is shown in parallel ($(R, h) = (30, 2)$~mm.). 




\end{document}